# Correlated spectrum of distant semiconductor qubits coupled by microwave photons

Baochuan Wang[1,2†], Ting Lin[1,2†], Haiou Li[1,2†], Sisi Gu[1,2], Mingbo Chen[1,2], Guangcan Guo[1,2], Hongwen Jiang[3], Xuedong Hu[4], Gang Cao[1,2*], Guoping Guo[1,2,5*]

1. *CAS Key Laboratory of Quantum Information, University of Science and Technology of China, Hefei 230026, China*
2. *CAS Center for Excellence in Quantum Information and Quantum Physics, University of Science and Technology of China, Hefei 230026, China*
3. *Department of Physics and Astronomy, University of California at Los Angeles, Los Angeles, CA 90095, USA*
4. *Department of Physics, University at Buffalo, SUNY, Buffalo, NY 14260-1500, USA.*
5. *Origin Quantum Computing Company Limited, Hefei 230026, China*

†These authors contributed equally to this work.
*Corresponding authors: gcao@ustc.edu.cn; gpguo@ustc.edu.cn

## ABSTRACT

We develop a new spectroscopic method to quickly and intuitively characterize the coupling of two microwave-photon-coupled semiconductor qubits via a high-impedance resonator. Highly distinctive and unique geometric patterns are revealed as we tune the qubit tunnel couplings relative to the frequency of the mediating photons. These patterns are in excellent agreement with a simulation using the Tavis-Cummings model, and allow us to readily identify different parameter regimes for both qubits in the detuning space. This method could potentially be an important component in the overall spectroscopic toolbox for quickly characterizing certain collective properties of multiple cavity QED coupled qubits.

## 1. Introduction

Semiconductor quantum dots are compatible with conventional manufacturing technology, and are a promising candidate for scalable quantum computing [1-3]. They have attracted significant interest [4,5] in systems based on GaAs [6-11], Si [12-18], and Ge [19,20]. However, as the number of qubits increases [21-24], the coupling of arbitrary pairs of distant qubits and the characterization of coupled qubits remain outstanding challenges.

In architectures based on circuit quantum electrodynamics (QED), photons in a microwave resonator are an effective medium in coupling distant quantum systems ranging from superconducting qubits to spin ensembles [25-27]. Photon-qubit interaction is also an efficient tool for quantum control and measurement of qubits [11,28-31]. In semiconductor systems, a series of pioneering works has explored coupling of quantum dots with a microwave resonator [32-41]. For example, strong coupling [25,26] has been demonstrated via observation of vacuum Rabi splitting in the frequency domain [42-47], and evidences of nonlocal qubit-qubit interactions in the strong coupling regime have been observed as well [48,49].

Existing experiments on circuit QED systems directly observe the energy spectrum of a system by sweeping the frequency of a probe microwave sent through the resonator. However, in the case of two or multiple qubits coupled to a cavity, with all near resonance, the total system spectrum and dynamics are determined by the interactions, and some states are not accessible through microwave (dark states). For questions such as whether the subsystems are all in resonance, the frequency-sweep is not a particularly sensitive probe, making the determination of such condition a laborious process.

Here we develop an alternative approach to study the collective properties of a semiconductor circuit QED architecture with multiple qubits. Specifically, we measure the microwave response of two qubits that are coupled to a single resonator at the cavity resonance frequency while scanning the detunings of both qubits, revealing the two-qubit spectrum from an angle different from the conventional approach. We systematically explore the evolution of the two-qubit spectrum as we vary the tunnel couplings for both qubits, and discover distinctive geometric patterns for each parameter regime in the detuning space. Furthermore, we find a one-to-one correspondence between these geometric features and the system parameters. Simulation using the Tavis-Cummings (TC) model [50,51] produces excellent agreement with our experimental findings. These results demonstrate that we can characterize properties of a circuit QED coupled hybrid system without sweeping the frequency of the applied microwave. And certain correlated properties such as two-qubit resonance between nonlocal semiconductor qubits can be characterized quickly and intuitively via pattern recognition. In other words, our spectroscopy method provides an alternative perspective on any two-qubit or multi-qubit cavity QED system, and can potentially speed up the process of characterizing parameters and preparing for certain collective configurations.

## 2. Results

### 2.1. Sample design

Fig. 1a shows our device of two gate-defined double quantum dots (DQDs) and a high-impedance ($Z_r \sim 1$ k$\Omega$ [42], which far exceeds the typical 50 $\Omega$ impedance of a traditional superconducting CPW resonator) $\lambda/4$ superconducting quantum interference device (SQUID) array resonator (red in Fig. 1a) fabricated on a GaAs/AlGaAs heterostructure. The left and right gate-defined DQDs encode two distant charge qubits, denoted as DQD$_1$ and DQD$_2$, with transition frequencies of $f_{a,i} = \omega_{a,i}/2\pi = \sqrt{\delta_i^2 + (2t_i)^2}$, respectively ($i = 1,2$). The interdot tunnel coupling $2t_{1(2)}$ can be tuned with gates 1MU (2MU) and 1MD (2MD). Gates 1BL (2BL) and 1BR (2BR) control the detuning $\delta_{1(2)}$ between the left dot and right dot of DQD$_{1(2)}$. To minimize crosstalk, we sweep the detuning by varying $V_{1BL(2BR)}$, which is far from the middle gates 1MU (2MU) and 1MD (2MD). Hence the tunnel coupling can be tuned independently while scanning the qubit detunings of both qubits.

The high-impedance resonator enables strong coupling between the qubits and the resonator. At the voltage antinode, the resonator extends two metal gates (gate 1PR and gate 2PL) to couple with the two DQDs. The resonator is also coupled to a drive line (cream in Fig. 1a) through an interdigital capacitor. The device can be represented by a circuit diagram shown in Fig. 1b.

## 2.2. Characterization of the resonator and individual DQDs

Fig. 1c shows the reflectance spectrum of the resonator without the qubits. According to the input-output theory [37,49], the microwave reflection of the resonator is given by $S_{11} = -1 + \frac{\kappa_{\text{ext}}}{i(\omega_{\text{r}} - \omega_{\text{p}}) + \kappa_{\text{tot}}/2}$, where $\omega_{\text{r}}$ and $\omega_{\text{p}}$ are the resonator and probe angular frequency. Fitting the data with this formula, we determine the resonator frequency at $f_{\text{r}} = \omega_{\text{r}}/2\pi = 6.48$ GHz, and its internal, external and total loss rate as $(\kappa_{\text{int}}, \kappa_{\text{ext}}, \kappa_{\text{tot}})/2\pi = (9.60, 25.46, 35.06)$ MHz.

Theoretically, our hybrid cavity QED system of two two-level artificial atoms interacting with a cavity can be described by the Tavis-Cummings (TC) model with the Hamiltonian [50]

$$H_{\text{TC}} = \omega_{\text{r}} a^+ a + \frac{1}{2}(\omega_{\text{a},1}\sigma_{\text{z},1} + \omega_{\text{a},2}\sigma_{\text{z},2}) + [g_1(\sigma_{+,1}a + a^+\sigma_{-,1}) + g_2(\sigma_{+,2}a + a^+\sigma_{-,2})], \quad (1)$$

where we have set $\hbar = 1$. The first two terms are for a free resonator and two independent charge qubits. The resonator frequency is $\omega_{\text{r}}/2\pi$, and $a^+(a)$ is the photon creation (annihilation) operator for the resonator mode. The charge qubit transition (operation) frequencies are $f_{\text{a},i} = \omega_{\text{a},i} /2\pi = \sqrt{\delta_i^2 + (2t_i)^2}$, where $\delta_i$ and $2t_i$ are the detuning and tunnel coupling, respectively. The Pauli matrices $\sigma_{\text{z},i}$ and $\sigma_{+/-,i}$ are defined in the charge qubit eigenbasis. The last term of the Hamiltonian gives the interactions between DQD$_i$ and the resonator. $g_i = g_i^{\max} \frac{2t_i}{\sqrt{(2t_i)^2 + \delta_i^2}}$ is the qubit-resonator coupling strength, with $g_i^{\max}$ its maximum value [37,49]. According to input-output theory, the microwave response of the cavity is modified by the two qubits through Hamiltonian (1), yielding a microwave reflection amplitude of

$$S_{11} = -1 + \frac{\kappa_{\text{ext}}}{i(\omega_r - \omega_p) + g_1\chi_1 + g_2\chi_2 + \kappa_{\text{tot}}/2}, \qquad (2)$$

where $\chi_i = \frac{g_i}{i(\omega_{a,i} - \omega_p) + \gamma_i}$ is the single-electron electrical susceptibility, and $\gamma_i$ is the decoherence rate of qubit$_i$ (see the Supplementary materials for details).

### 2.3. Correlated spectra in different coupling regimes

We first examine the coupling between a single qubit (DQD$_1$) and the resonator. Here DQD$_2$ is detuned far away from the resonator frequency and can be neglected. By fixing the probe frequency at the resonator frequency $f_p = f_r = 6.48$ GHz, Eq. (2) can be simplified to $S_{11} = -1 + \frac{\kappa_{\text{ext}}}{g_1\chi_1 + \kappa_{\text{tot}}/2}$, and we measure charge stability from the microwave response. Fig. 2a shows the scenario when $2t_1 = 6.55$ GHz $> f_r$. The qubit transition frequency $f_{a,1}$ is always larger than the resonator frequency $f_r = f_p$ no matter what $\delta_1$ is. In particular, $f_{a,1} - f_p$ is minimum at $\delta_1 = 0$, thus maximizing $|S_{11}|$ at $\delta_1 = 0$ and resulting in a single transition line in the charge stability diagram. Hence, in Fig. 2c there exists only one peak when we cut a line along the detuning of qubit 1 (white arrow in Fig. 2a). In Fig. 2b, where $2t_1 = 5.9$ GHz $< f_r$, the qubit can become resonant with the resonator. Specifically, $f_{a,1} = f_r$ when $\delta_1 = \pm\sqrt{f_r^2 - (2t_1)^2}$, leading to two maxima in $|g_1\chi_1|$. Consequently, there are two transition lines in Fig. 2b, and two separate peaks in Fig. 2d. The qubit-resonator resonance condition $\omega_{a,1} = \omega_r$ can be established at the point where the single Lorentzian peak split into two. We can perform the same study of coupling between DQD$_2$ and the resonator, and observe similar results.

We now study our full system of two qubits coupled to the resonator. Based on our philosophy of reducing effect of the resonator, we again fix the probe frequency on resonance with the resonator, $f_\mathrm{p} = f_\mathrm{r}$. An additional benefit is that all parameters are now controlled electrically. Our main goal is to investigate the nonlocal coherent coupling and two-qubit spectra by measuring microwave reflection as a function of the detuning of each qubit, with

$$S_{11}(\delta_1, \delta_2) = -1 + \frac{\kappa_\mathrm{ext}}{g_1\chi_1 + g_2\chi_2 + \kappa_\mathrm{tot}/2}. \qquad (3)$$

While Eq. (3) takes on a seemingly simple form, the number of underlying parameters is still quite large. Here we parametrize the system by the relative strength of the tunnel coupling rate $t_i$ of DQD$_i$ and the resonator frequency $f_\mathrm{r}$, so that there are nine different coupling regimes. To prevent repetition, and without loss of generality, we now focus on the behavior and evolution of the two-qubit spectra of four characteristic coupling regimes (Table 1):

(i) For $(2t_1, 2t_2) = (6.68, 6.59)$ GHz $> f_\mathrm{r}$, as shown in Fig. 3a, a red crossed pattern with a deep red center intuitively shows that $|S_{11}|$ increases as $|\delta_i|$ decreases and is maximum at $|\delta_1| = |\delta_2| = 0$. In this regime both qubits are weakly coupled with the resonator, and the qubits' contributions can be regarded as a shift of the resonator frequency $\Delta\omega_\mathrm{r} = \mathrm{Im}\{g_1\chi_1 + g_2\chi_2\}$, resulting in the change in $|S_{11}|$ around $|\delta_1| = |\delta_2| = 0$ being almost a linear superposition of responses from individual qubits [37,52].

(ii) For $(2t_1, 2t_2) = (6.68, 6.48)$ GHz, $2t_2 = f_\mathrm{r} < 2t_1$, as shown in Fig. 3b, the single peak in the central area of Fig. 3a is split into two peaks that lie symmetrically on either side of $\delta_2 = 0$. In the central area where $g_2\chi_2 \gg g_1\chi_1$, the contribution of qubit$_2$ is much larger than that of qubit$_1$ and plays the leading role in the microwave response. However, qubit$_1$ does modify the signal, leading to the two peaks being extended out, as illustrated in the center of Fig. 3b.

(iii) For $2t_1 = 2t_2 = f_\mathrm{r} = 6.48$ GHz, the qubits are operating in the strong coupling

limit. From Eq. (3), $g_i\chi_i$ will dominate the microwave response when $g_i \sim g_i^{\max}$, especially if $g_i\chi_i$ is greater than $\kappa_{\text{tot}}/2$. This is the limit when the two qubits and the resonator are completely mixed, which leads to an interesting pattern in Fig. 3c. In particular, in the central area, the enhanced reflection extends outward along the diagonal lines, where the two qubits become detuned from the resonator but are still resonant with each other with $f_{\text{a},1} = f_{\text{a},2}$, resulting in a distinct X-like pattern in the reflection signal. The two qubits are the most strongly coupled via the resonator along the diagonals, with an effective coupling strength $g_{\text{eff}} = \sqrt{g_1^2 + g_2^2}$. Consequently, the enhanced reflection signal is detected in an X-like pattern extending along both diagonal directions.

(iv) Finally, with the DQD's tunnel coupling tuned away from $2t_2 = 6.48$ GHz to $2t_2 = 6.42$ GHz, which means that $2t_2 < 2t_1 = f_{\text{r}}$, the X-like pattern is stretched in the horizontal ($\delta_2$) direction in Fig. 3d. The single enhanced peak in the central area of Fig. 3c splits into two peaks at the corresponding detuning points $\delta_2 = \pm\sqrt{f_{\text{r}}^2 - (2t_2)^2}$, where $f_{\text{a},2} = f_{\text{r}}$ in Fig. 3d. The additional two bright valley spots, caused by the cancellation between the two qubits (see the Supplementary materials for details), appear on the sides of the X-axis around $\delta_2 = 0$.

The top panels in Fig. 3 show that by scanning the gate voltages for the two qubits, the geometrical patterns in different tunnel coupling regimes exhibit unique features that intuitively provide information about the frequency relationships among the resonator and the two qubits. To confirm this point, we simulate these correlated spectra using averaged typical parameters of $(g_1^{\max}, g_2^{\max}, \gamma_1, \gamma_2)/2\pi = (86, 85, 22, 23)$ MHz from Refs. [42,48] with their corresponding tunnel coupling rates. The simulated two-qubit spectra, shown in Figs. 3e–h, qualitatively profile our experimental evolution very well. However, the color scale of the patterns does not match our experimental results quantitatively, since the input parameters are estimated typical values instead of the real values in our hybrid system. The most intriguing question here is whether these parameters can change the geometrical patterns and whether we can extract our system parameters from the experimentally established correlation spectrum.

To understand the dependence of the spectrum patterns on the parameters $(g_1^{\mathrm{max}}, g_2^{\mathrm{max}}, \gamma_1, \gamma_2)/2\pi$ quantitatively, we focus on the geometric features of the X-like pattern in Fig. 3c, where each qubit is at or near its maximum coupling strength $g_i^{\mathrm{max}}$, and systematically investigate its evolution theoretically. Since both qubits are fabricated on the same wafer, confined with the same structure and operated in the same environment, we take $\gamma_1 = \gamma_2 = \gamma$ for simplicity without loss of generality. Here, we use a few typical parameters to characterize the features of the X-like pattern. The first parameter is the ratio $\delta_1^{\mathrm{FHWM}}/\delta_2^{\mathrm{FHWM}}$, where $\delta_{1(2)}^{\mathrm{FHWM}}$ is the full width at half maximum (FHWM) along the Y(X)-axis, as shown in Fig. 4a. Fig. 4b shows the calculated results of $\log_{10}\left(\delta_1^{\mathrm{FHWM}}/\delta_2^{\mathrm{FHWM}}\right)$ as a function of $g_1^{\mathrm{max}}/2\pi$ and $g_2^{\mathrm{max}}/2\pi$. From the contour line in Fig. 4b, it is clearly seen that $\log_{10}\left(\delta_1^{\mathrm{FHWM}}/\delta_2^{\mathrm{FHWM}}\right)$ is almost linearly related to the ratio of $g_1^{\mathrm{max}}/g^{\mathrm{max}}$. Combined with our experimental data $\log_{10}\left(\delta_1^{\mathrm{FHWM}}/\delta_2^{\mathrm{FHWM}}\right) = 0.11$ extracted from Fig. 3c, we can estimate $g_1^{\mathrm{max}} \approx g_2^{\mathrm{max}} = g$.

We then take a line cut along the diagonal $\delta_1 = \delta_2$ in Fig. 4a and use $\delta_{12}^{\mathrm{FHWM}}$ and $|S_{11}|^{\mathrm{max}}$ to denote the FHWM and maximum value of the curve, respectively. The calculated result of the FHWM as a function of $g/2\pi$ and $\gamma/2\pi$ shown in Fig. 4c indicates that $\delta_{12}^{\mathrm{FHWM}}$ is almost completely determined by the coupling rate $g/2\pi$ and is insensitive to the decoherence rate $\gamma/2\pi$. This is because around the half maximum of the curve along the off-diagonal cut $\delta_1 = \delta_2$ in Fig. 4c, $\omega_{\mathrm{a},i} - \omega_{\mathrm{p}} \gg \gamma_i$, so that the qubits' contributions $g_i\chi_i = \dfrac{g_i^2}{i(\omega_{\mathrm{a},i}-\omega_{\mathrm{p}})+\gamma_i}$ in Eq. (1) are only weakly affected by variation in $\gamma_i$. Therefore, using $\delta_{12}^{\mathrm{FHWM}} = 7.08$ GHz extracted from the experimental data in Fig. 3c, we obtain $g/2\pi = 80$ MHz. Fig. 4d presents the evolution of $|S_{11}|^{\mathrm{max}}$ as a function of $g/2\pi$ and $\gamma/2\pi$. Combining the estimated $g/2\pi = 80$ MHz with the experimental value $|S_{11}|^{\mathrm{max}} = 0.90$, $\gamma/2\pi = 55$ MHz can then be determined  (see

Fig. S7 online for details).

The estimated parameters $(g_1^{max}, g_2^{max}, \gamma_1, \gamma_2)/2\pi = (80, 80, 55, 55)$ MHz are consistent with our vacuum Rabi splitting experiment (see Fig. S6 online for details). Using the estimated parameters, two-qubit spectra for different parameter regimes have been simulated in Figs. 3i–l. The calculated spectra not only describe our experimentally found pattern evolution qualitatively, but also reproduce the color scale of the patterns quantitatively.

## 3. Conclusion

In conclusion, we have investigated the coupling of two microwave-photon-coupled qubits with a new spectroscopic approach. Compared to the conventional approach, we directly sweep the qubit frequencies, and find that the corresponding measured microwave response displays distinctive geometric patterns for different parameter regimes of the coupled qubit-resonator system in the detuning space. Pattern recognition can then allow quick and intuitive characterization of interactions between any pair of distant qubits.

Our method provides a new perspective into the huge parameter space of a multi-qubit cavity QED coupled system, and is sensitive to collective properties such as whether the subsystems are in resonance, which we have discussed in Section S7 in the Supplemental Material. Most importantly, this example clearly demonstrates that for an arbitrary multi-parameter system, examining the system from multiple angles (different cuts through the high-dimensional parameter space) could reveal information that is otherwise hidden. For example, in our system sweeping the tunnel couplings or the coupling strengths of qubits could prove to be insightful as well, and will be explored in the future.

Our study is based on the Tavis-Cumming model and the corresponding input-output theory, and our results show that the semiconductor-resonator hybrid system can be perfectly described by this general cavity QED theory. As such our method is a general approach and could be adopted in other multi-qubit cavity QED systems as well. For example, recent works demonstrating strong spin-photon coupling employ a spin-charge conversion scheme, where our spectroscopy method can be easily applied (see Section S9 in the Supplemental materials for details). Meanwhile, machine learning techniques may be adopted to improve pattern recognition from noisy data.

In short, our approach is a clear illustration that looking into a general multi-qubit system from different angles could reveal distinctive information of such a multi-variable system, and could form an integral part of a multi-pronged and more efficient query into a multi-qubit cavity QED system.

**Conflict of interest**

The authors declare that they have no conflict of interest.

**Acknowledgments**

This work was supported by the National Key Research and Development Program of China (2016YFA0301700), the National Natural Science Foundation of China (61922074, 11674300, 61674132, 11625419 and 11804327), the Strategic Priority Research Program of the CAS (XDB24030601), and the Anhui initiative in Quantum Information Technologies (AHY080000). Hongwen Jiang and Xuedong Hu acknowledge financial support by U.S. ARO through Grant No. W911NF1410346 and No. W911NF1710257, respectively. This work was partially carried out at the University of Science and Technology of China Center for Micro and Nanoscale Research and Fabrication.

**Author contributions**

Baochuan Wang, Haiou Li, and Sisi Gu fabricated the sample. Ting Lin, Gang Cao, Baochuan Wang, and Haiou Li performed the measurements and ran simulations. Ting

Lin, Gang Cao, Mingbo Chen, Xuedong Hu, Hongwen Jiang and Guoping Guo analyzed the date. Guangcan Guo and Guoping Guo advised on experiments and data analysis. Guoping Guo and Cao Gang supervised the experiments. All authors contributed to write the paper processor in silicon. Nature 2018;555:633.

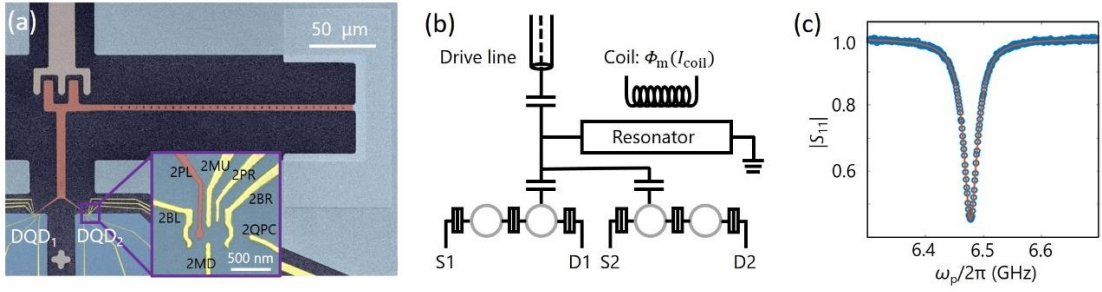

FIG. 1 (Color online) (a) False-color electron micrograph of our device. The resonator (red) is capacitively coupled with the drive line (cream) and two DQDs. A coil is mounted above the device to adjust the resonator frequency by changing the flux current $I_{\text{coil}}$. The inset shows an enlarged view of DQD$_2$. (b) A simplified circuit diagram of (a). The probe microwave is applied to the hybrid system through the drive line. The input and reflected microwave are separated by a circulator (not shown). $\phi_{\text{m}}$ is the applied magnetic flux bias. (c) Measured resonator reflectance (blue circles) as a function of the probe frequency $\omega_{\text{p}}/2\pi$.

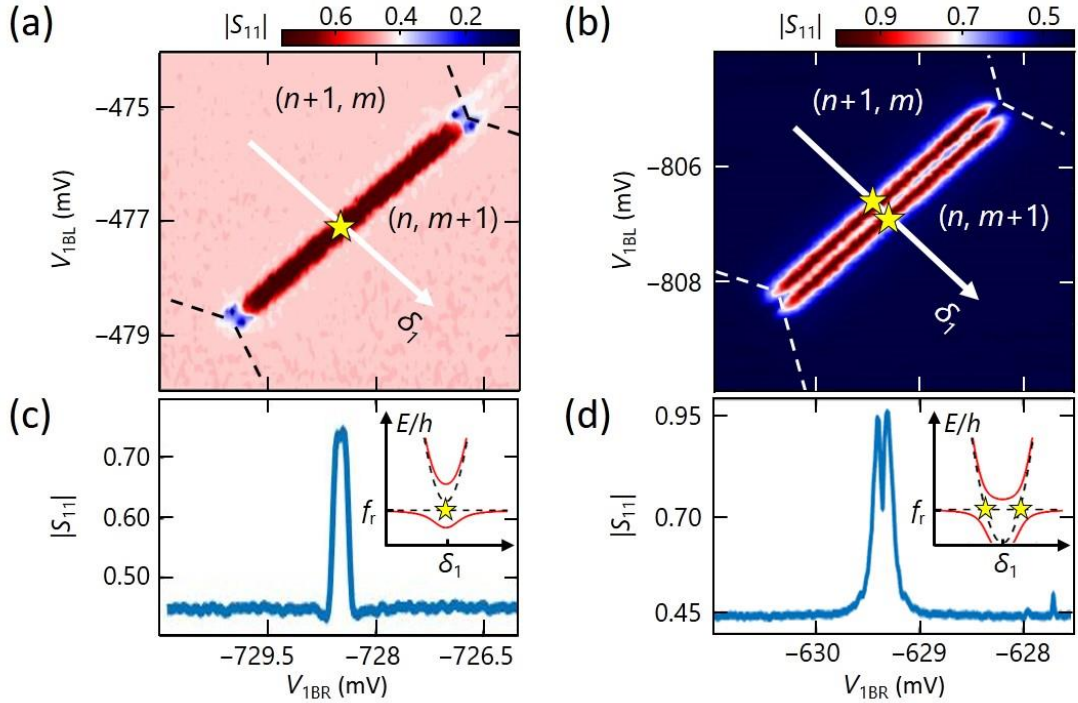

Fig 2 (Color online) Charge stability diagrams of DQD$_1$ extracted from $|S_{11}|$ with the probe frequency fixed at $f_p = f_r$. The numbers $(n, m)$ denote the charge number in the left and right dots, respectively. Measured $|S_{11}|$ as a function of $V_{1BL}$ and $V_{1BR}$ at (a) $2t_1 = 6.55$ GHz $> f_r$ and (b) $2t_1 = 5.90$ GHz $< f_r$. (c, d) are the cut lines along the white arrows in (a) and (b), respectively. The stars are used to illustrate the position of the peaks. Insets in (c) and (d) are the energy level diagrams of DQD$_1$ and the resonator. The solid (dashed) lines correspond to the situation when DQD$_1$ is (is not) coupled with the resonator.

Table 1 Geometric features at different coupling regimes

| Figs. 3a, e and i | Figs. 3b, f and j | Figs. 3c, g and k | Figs. 3d, h and l |
|---|---|---|---|
| $2t_1, 2t_2 > f_r$ | $2t_2 = f_r < 2t_1$ | $2t_1, 2t_2 = f_r$ | $2t_2 < f_r = 2t_1$ |
| Linear superposition | Two separated peaks | X-like pattern | Two separated spots |

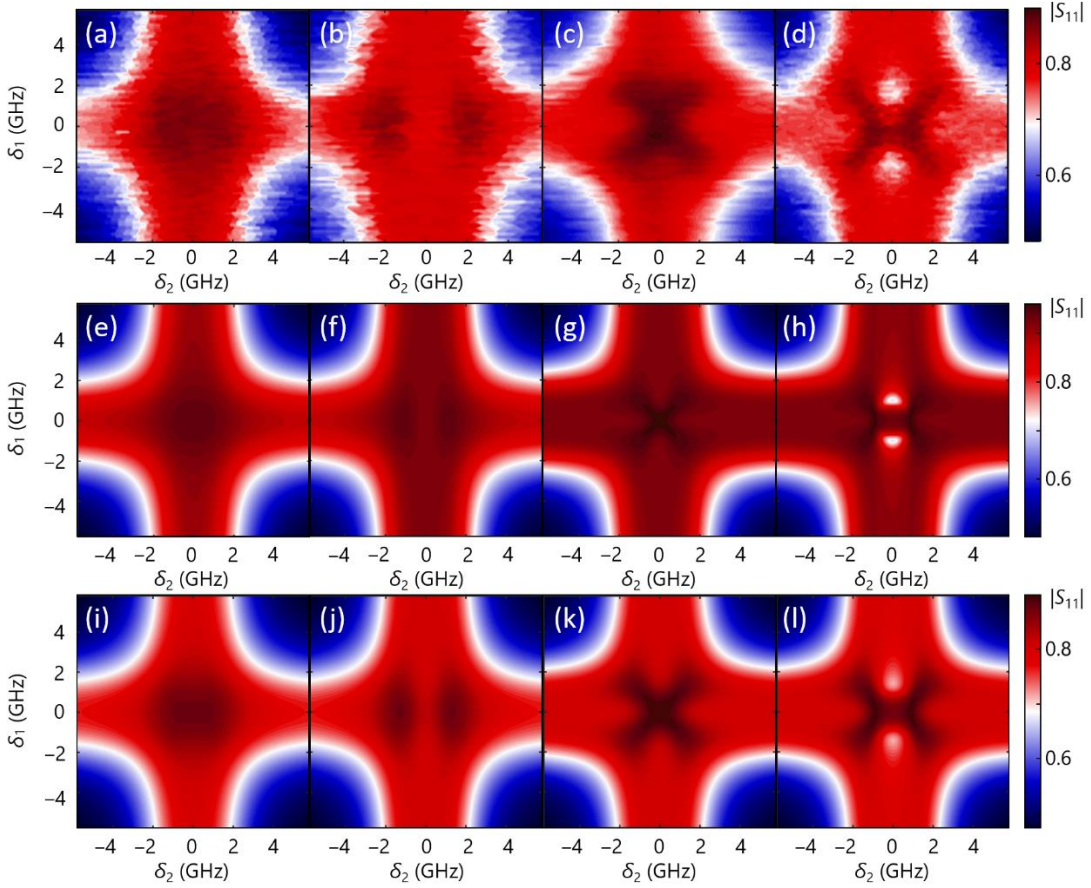

FIG. 3 (Color online) Measured two-qubit spectra as a function of $\delta_1$ and $\delta_2$ at $f_p = f_r = 6.48$ GHz, with (a) $(2t_1, 2t_2) = (6.68, 6.59)$ GHz, (b) $(2t_1, 2t_2) = (6.68, 6.48)$ GHz, (c) $(2t_1, 2t_2) = (6.48, 6.48)$ GHz, and (d) $(2t_1, 2t_2) = (6.48, 6.42)$ GHz. Calculations corresponding to (a-d) with typical parameters $(g_1^{max}, g_2^{max}, \gamma_1, \gamma_2)/2\pi = (86, 85, 22, 23)$ MHz and extracted parameters $(g_1^{max}, g_2^{max}, \gamma_1, \gamma_2)/2\pi = (80, 80, 55, 55)$ MHz are shown in (e–h) and (i–l) respectively.

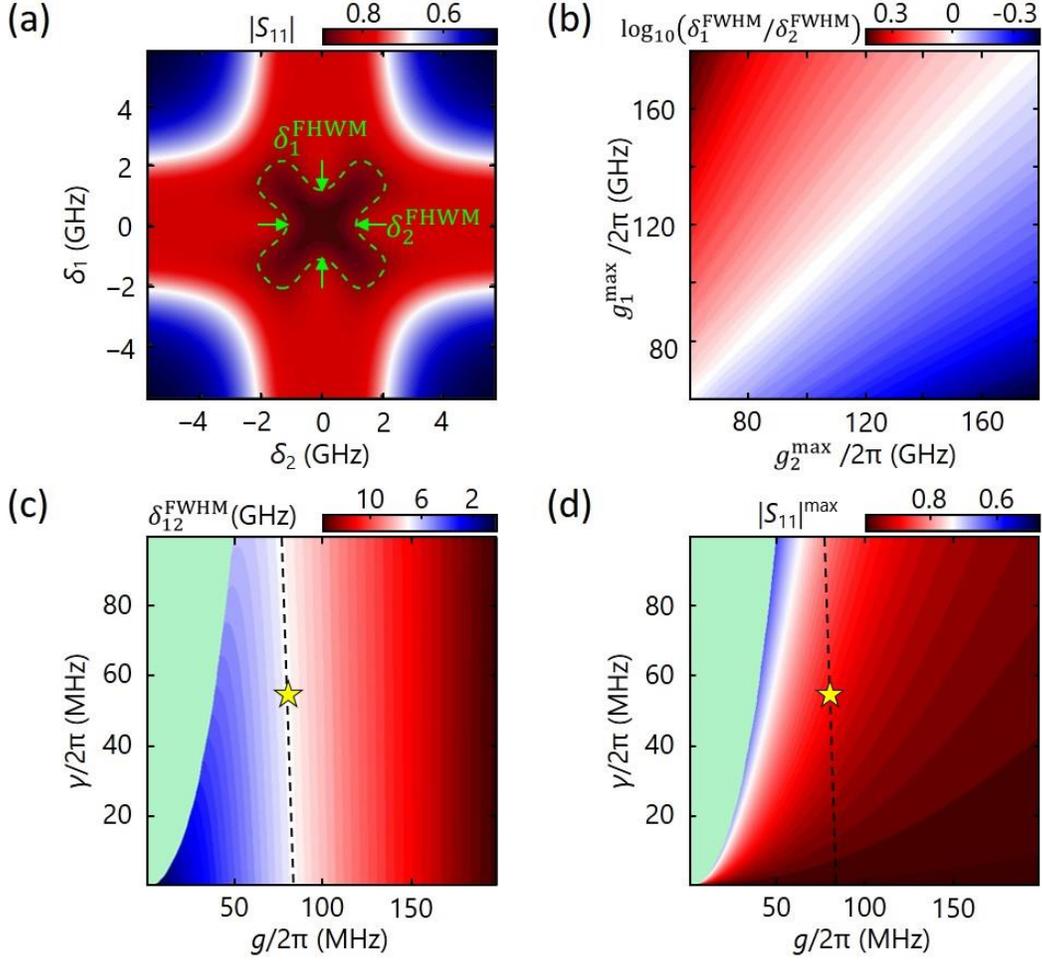

FIG. 4 (Color online) (a) Simulated two-qubit spectrum corresponding to Fig. 3c with $(g_1^{\mathrm{max}}, g_2^{\mathrm{max}}, \gamma_1, \gamma_2)/2\pi = (80, 80, 55, 55)$ MHz. (b) $\delta_1^{\mathrm{FWHM}}/\delta_2^{\mathrm{FWHM}}$ as a function of $g_1^{\mathrm{max}}/2\pi$ and $g_2^{\mathrm{max}}/2\pi$. (c, d) are $\delta_{12}^{\mathrm{FWHM}}$ and $|S_{11}|^{\mathrm{max}}$ as a function of $g = g_1^{\mathrm{max}} = g_2^{\mathrm{max}}$ and $\gamma = \gamma_1 = \gamma_2$. The yellow stars refer to $(g_1^{\mathrm{max}}, g_2^{\mathrm{max}}, \gamma_1, \gamma_2)/2\pi = (80, 80, 55, 55)$ MHz. The green regions refer to the regimes in which the X-like pattern disappears where both qubits are weakly coupled with the resonator.

# Supplemental Material: Correlated spectrum of distant semiconductor qubits coupled by microwave photons

## Section S1. **The microwave response in different regimes**

According to input-output theory [1,2], the microwave response of a cavity coupled to two qubits is modulated by Hamiltonian (Eq. (1)), yielding a microwave reflection amplitude of

$$S_{11}(\delta_1, \delta_2) = -1 + \frac{\kappa_{\text{ext}}}{i(\omega_{\text{r}} - \omega_{\text{p}}) + g_1\chi_1 + g_2\chi_2 + \kappa_{\text{tot}}/2}, \qquad \text{(S1)}$$

In the main text, we fix the probe frequency on resonance with the resonator, $\omega_{\text{p}} = \omega_{\text{r}}$, then investigate the nonlocal coherent coupling by measuring and analyzing the correlated two-qubit spectra as a function of the detuning of each qubit. By choosing the probe frequency to be resonant with the resonator, we have removed the extraneous contribution of probe-resonator mismatch from the microwave response in Eq. (S1), and singled out the qubit contributions $g_i\chi_i$. Therefore, Eq. (S1) can now be simplified as:

$$S_{11}(\delta_1, \delta_2) = -1 + \frac{\kappa_{\text{ext}}}{g_1\chi_1 + g_2\chi_2 + \kappa_{\text{tot}}/2}, \qquad \text{(S2)}$$

Here, we study the evolution of the correlated spectra between different coupling regimes defined in terms of the tunnel coupling $(2t_1, 2t_2)$ of the two qubits. Our description of the qualitative behaviors is based on Eq. (S2). The regimes of interest are:

(i) $(2t_1, 2t_2) = (6.68, 6.59)$ GHz $> f_r$. Qubit contributions $g_i\chi_i$ in Eq. (S2) are small compared with the microwave loss $\kappa_{\text{tot}}/2$ via the resonator in this regime. Expanding Eq. (S2) to the first order of $g_i\chi_i$, we get $S_{11}(\delta_1, \delta_2) \approx -1 + \frac{2\kappa_{\text{ext}}}{\kappa_{\text{tot}}}(1 - \frac{g_1\chi_1 + g_2\chi_2}{\kappa_{\text{tot}}/2})$. When $\Delta_i \gg \gamma_i$ ($\Delta_i = \omega_i - \omega_p$), the change of microwave reflectance $\Delta|S_{11}|$ is proportional to the sum of $|g_1\chi_1|$ and $|g_2\chi_2|$, as shown in Fig. 3a.

(ii) $(2t_1, 2t_2) = (6.68, 6.48)$ GHz. In this regime, the contribution of qubit₂ is much larger than that of qubit₁ and plays the leading role in the microwave response. However, around $\Delta_2 = 0$, $g_1\chi_1 \approx -ig_1^2/\Delta_1$ provides the imaginary part (here, to simplify our discussion, we assume that $\Delta_1$ is much larger than $\gamma_1$), and the Eq. (S2) can be written as $S_{11} \approx -1 + \frac{\kappa_{\text{ext}}}{(g_2\chi_2 + \kappa_{\text{tot}}/2)e^{i\theta}}$, where $\theta = -\arctan(\frac{g_1^2/\Delta_1}{g_2\chi_2 + \kappa_{\text{tot}}/2})$ is the factor to describe the asymmetry caused by sweeping the detuning of qubit₁. In Fig. 3b, with qubit₁ far detuned from both qubit₂ and the resonator, $\theta \approx 0$ and the microwave reflectance $S_{11} = -1 + \frac{\kappa_{\text{ext}}}{g_2\chi_2 + \kappa_{\text{tot}}/2}$ maximizes at $\Delta_2 = 0$ ($\delta_2 = 0$). As $|\delta_1|$ decreases, the increased $|\theta|$ modifies the signal, resulting in the enhancement of the peaks' value at larger $\Delta_2$ ($|\delta_2|$), i.e, the separated peaks extended out from $|\delta_2| = 0$.

(iii) $2t_1 = 2t_2 = f_r = 6.48$ GHz. In the central area, $g_1\chi_1$ and $g_2\chi_2$ are of the same

order of magnitude. When $\Delta_1(|\delta_1|)$ is detuned from 0, $g_1\chi_1 = \frac{g_1^2}{i\Delta_1+\gamma_1}$ provides the imaginary part which is 0 at $\Delta_2(|\delta_2|) = 0$ and increases within $\Delta_1 < \gamma_1$. Thus the separated peaks at the cut line along the detuning of qubit$_2$ arise and extend out at the cut lines as $\Delta_1(|\delta_1|)$ increases. Likewise, we observe the separated peaks at the cut lines along the detuning of qubit$_1$.

(iv) $2t_2 = 6.42$ GHz $< f_r = 2t_1 = 6.48$ GHz. If we only consider the interaction between qubit$_1$ and the resonator and measure the microwave reflectance $|S_{11}|$ as a function of $\Delta_1(\delta_1)$, there would exist only one peak. The qubit part $|g_1\chi_1| = \frac{g_1^2}{\sqrt{\gamma_1^2+\Delta_1^2}}$ maximizes at $\Delta_1 = 0$ ($\delta_1 = 0$) and decrease monotonically as $|\Delta_1|$ ($|\delta_1|$) increasing. However, differ from Fig. 2a, in this working regime, qubit$_2$'s inter-dot tunneling rate $2t_2$ is close to $f_r$ and the coupling between qubit$_2$ and the resonator provides an imaginary part which has the opposite sign compared with the imaginary part induced by qubit$_1$. Thus, the cancellation between two qubits is conspicuous around $(\delta_1, \delta_2) \sim (1.5, 0)$ GHz in Fig. 3d.

## Section S2. **Crosstalk between the two qubits**

In our sample structure, gates 1PR and 2PL are both connected to the center conductor of the SQUID array resonator. In order to determine the influence of the potential direct capacitive coupling between the two qubits, we measure the crosstalk of our hybrid system. Since gates 1PR and 2PL are floated without bias voltages, here we scan voltages on gate 1BR (close to gate 1PR) and gate 2BL (close to gate 2PL) to detect possibly larger crosstalk because of the proximity of the gates. We also scan voltages on gate 1BL (far from gate 1PR) and gate 2BR (far from gate 2PL) to look for evidence of smaller crosstalk.

In Fig. S1a, microwave response is plotted as a function of $V_{2BL}$ and $V_{1BR}$, where $V_{1BR}$ is scanned across the inter-dot transition line of qubit$_1$. Two green dashed lines in Fig. S1a indicate the degenerate points between qubit$_1$ and the resonator, since $2t_1 < f_r$. The slope of the dashed lines $k_1 \sim -1/215$ describes the capacitive relationship of the two nonlocal qubits. While the slope is non-vanishing, it is still quite small in magnitude. In Fig. S1b, the capacitive relationship between gates 2BR and 1BL is detected in the same way. Here the almost horizontal features with the slope of the dashed lines $k_2 \sim 0$ clearly show that these two gates have minimal crosstalk, as expected. In our experiments in Fig. 3, the central regions of the correlated spectra are within the range of $|\delta_i| < 4$ GHz, with the corresponding shift of the other qubit being $|\delta_i| \cdot \max(|k_1|, |k_2|) < 19$ MHz, or less than 1%. Hence, the effect of crosstalk between the two nonlocal qubits is negligible in our experiments.

When direct dipole coupling between the two qubits is finite, the transition lines for each qubit would be shifted [3]. However, in our experiment these were not observed. Therefore, we believe the direct dipole coupling is negligible.

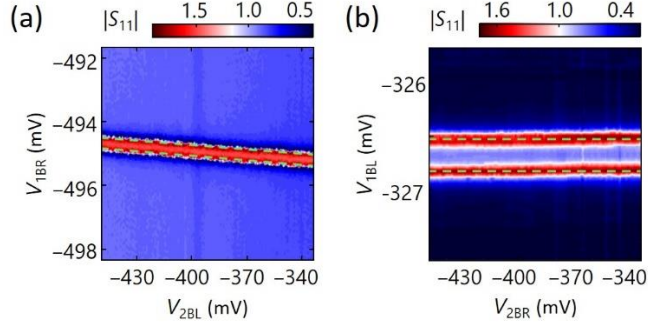

**FIG. S1.** Voltage crosstalk between nonlocal two qubits. (a)  Measured $|S_{11}|$ as a function of $V_{2BL}$ and $V_{1BR}$. (b) Measured $|S_{11}|$ as a function of $V_{2BR}$ and $V_{1BL}$. (In (a) and (b), qubit$_1$ has the same tunnel coupling rate $2t_1$).

### Section S3. **Crosstalk between the surface metal gates and the resonator**

As discussed in Section S2, crosstalk between the two qubits can be ignored. Here we show that the cavity's frequency shift caused by changing surface metal gates' voltages can also be neglected. Figure S2 below shows the resonator reflectance as we vary $V_{2BL}$. The resonator frequencies are $\omega_r/2\pi = 6.513$ GHz at $V_{2BL} = -350$ mV and 6.514 GHz at $V_{2BL} = -400$ mV (these data are from a different cooldown than that in the main text, so that the resonator frequency is slightly different).

In our experiments reported in Fig. 3, the central regions are within the range of $|\delta_2| < 4$ GHz, which corresponds to a change of $V_{2BL}$ of less than 0.44 mV. Based on our finding from Fig. S2, such a small voltage change in $V_{2BL}$ would only lead to a shift of resonator frequency of $\Delta\omega_r/2\pi < 17$ kHz, which is much smaller than any other energy scale in the system. We can thus safely neglect the resonator frequency change in our experiments.

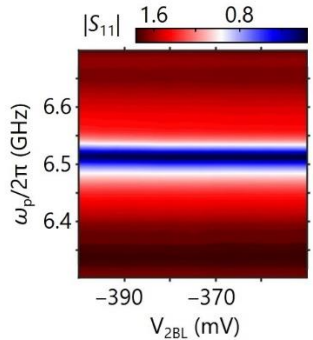

**FIG. S2.** Measured $|S_{11}|$ as a function of $V_{2BL}$ and $\omega_p/2\pi$.

### Section S4. **Coupling strength dependence in Fig. 3a**

For $(2t_1, 2t_2) = (6.68, 6.59)$ GHz $> f_r$, both qubits far are detuned from $f_r$, so that qubit contributions $g_i \chi_i$ in Eq. (S2) are small compared with the resonator loss $\kappa_{tot}/2$ in this regime. The interactions between the resonator and the qubits causes an increase in the reflectance at $\delta_1 = \delta_2 = 0$. Figure S3 shows the change of microwave reflectance $\Delta|S_{11}|$ as a function of $g_1^{\mathrm{max}}$ and $g_2^{\mathrm{max}}$. In the weak coupling regime where $g_1^{\mathrm{max}}, g_2^{\mathrm{max}} \ll \gamma = 50$ MHz, the calculated result is consistent with the simplified equation $S_{11}(\delta_1, \delta_2) \approx -1 + \frac{2\kappa_{\mathrm{ext}}}{\kappa_{\mathrm{tot}}}(1 - \frac{g_1\chi_1 + g_2\chi_2}{\kappa_{\mathrm{tot}}/2})$, leading to the change of microwave reflectance $|\Delta S_{11}|$ which is proportional to the sum of $|g_1\chi_1|$ and $|g_2\chi_2|$.

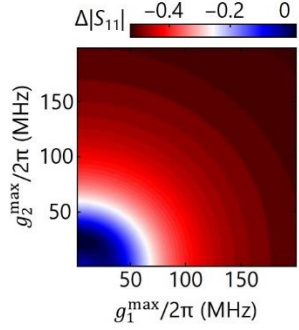

**FIG. S3.** Microwave response of two qubits at $\delta_1 = \delta_2 = 0$. $\Delta|S_{11}|$ as function of $g_1^{\mathrm{max}}/2\pi$ and $g_2^{\mathrm{max}}/2\pi$.

## Section S5. **Geometric features of the separated peaks in Fig. 3b**

In Fig. 3b of the main text, with $2t_2 = f_r = 6.48$ GHz and the inter-dot tunnel coupling of DQD$_1$ tuned to $2t_1 = 6.68$ GHz, two peaks symmetrically lie on the two sides of $\delta_1 = 0$. In the working regime of ($|g_2\chi_2| \gg |g_1\chi_1|, \kappa_{\mathrm{tot}}$), the contribution of qubit$_2$ plays the leading role in the microwave response. However, with the modification of the imaginary part of $g_1\chi_1$, the maximum value of microwave reflectance does not appear at $\delta_1 = \delta_2 = 0$, but locates at two different positions on the vertical-axis ($\delta_1 = 0$). Here we characterize the features of these separated peaks with $\delta_2^{\mathrm{d}}$ and $|S_{11}|^{\mathrm{max}}$, where $\delta_2^{\mathrm{d}}$ is the distance between the two peaks along $\delta_1 = 0$ and $|S_{11}|^{\mathrm{max}}$ is the height of these peaks.

First, we investigate the evolutions of geometric features as a function of $(g, \gamma)$. The calculated result in Fig. S4a predicts that the distance between two peaks $\delta_2^{\mathrm{d}}$ increases as $\gamma$, $g$ increase, while Fig. S4b shows that $|S_{11}|^{\mathrm{max}}$ is enhanced with the increase of coupling rate $g$ or the decrease of decoherence rate $\gamma$. We then examine the dependence of the geometric features on the changing $(g_1^{\mathrm{max}}, g_2^{\mathrm{max}})$. In Fig. S4c, the distance between the two peaks $\delta_2^{\mathrm{d}}$ increases as $g_1$ increases. On the other hand, the microwave response is affected by both qubits, and the relative weight of the contribution from qubit$_1$ is lessened as $g_2$ increases. In this scenario, the two peaks approach each other and degenerate into a single one. Figure S4d shows that $|S_{11}|^{\mathrm{max}}$, which contains contribution from both $g_1\chi_1$ and $g_2\chi_2$, is enhanced with the increase of $g_1$ and $g_2$, as one would expect from Eq. (3).

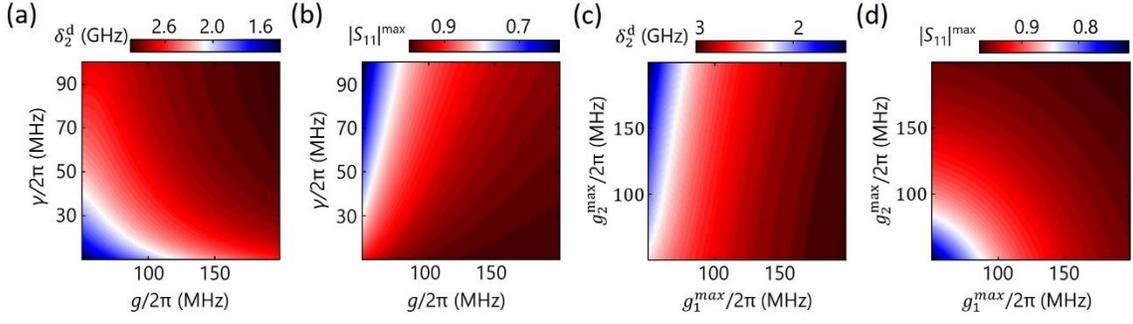

**FIG. S4.** Geometric features of the separated peaks. (a, b) $\delta_2^d$ and $|S_{11}|^{max}$ as function of coherence rate $g = g_1^{max} = g_2^{max}$ and decoherence rate $\gamma = \gamma_1 = \gamma_2$, respectively. (c, d) $\delta_2^d$ and $|S_{11}|^{max}$ as a function of $g_1^{max}$ and $g_2^{max}$.

## Section S6. **Geometric features of the bright spots in Fig. 3d**

In Fig. 3d of the main text, with $2t_1 = f_r$ and the inter-dot tunnel coupling rate $2t_2$ of DQD$_2$ tuned from $2t_2 = 6.48$ GHz to $6.42$ GHz, two bright spots appear on the two sides of the horizontal axis along $\delta_1 = 0$. These spots represent a dip in the reflectance, which is caused by the opposite sign of the imaginary parts of $g_1\chi_1$ and $g_2\chi_2$ in Eq. (S2) when $|\delta_2| < \sqrt{f_r^2 - (2t_2)^2}$. , This opposite phase shift means that the influences of the two qubits cancel out with each other, resulting in the two dips of microwave reflectance. Here we characterize the two bright spots with $\delta_1^v$ and $C$, where $\delta_1^v$ is the separation between the two dips along $\delta_2 = 0$, and $C(g, \gamma, t_2') = \frac{|S_{11}(\delta_1^v/2, \delta_2=0; t_1, t_2=f_r)| - |S_{11}(\delta_1^v/2, \delta_2=0; t_1=f_r, t_2=t_2')|}{|S_{11}(\delta_1^v/2, \delta_2=0; t_1, t_2=f_r)|}$ is basically the contrast of the dips.

Here we provide additional information on the evolutions of these bright spots as a function of different $(g, \gamma)$ using Eq. (3). The calculated results in Fig. S5a show that the positions of the bright spots are nearly independent from the coupling rate $g$, and are pushed apart by increasing $\gamma$. Figure S5b shows that the bright spots should become more conspicuous as $\gamma, g$ decrease. We also calculate the dependence of the geometric features on different $(g, 2t_2)$. In Fig. S5c, when $2t_2$ increases and approaches $f_r = 6.48$ GHz, $\delta_1^v$ increases quickly and eventually diverges, but is nearly independent from $g$. Figure S5d shows that the bright spots should fade as both $g, 2t_2$ increase. In short, as $2t_2$ approaches the resonator frequency from below, the bright spots gradually fade and move away from the center region of the correlated spectrum in Fig. 3d. When $2t_2$ reaches $6.48$ GHz, the bright spots would disappear and an X-like pattern develops, as shown in Fig. 3c. Therefore, the presence of these bright spots, particularly their positions and contrast, can be used as indicators on whether $2t_2$ is approaching $f_r$ from below. If $2t_2 > f_r = 2t_1$, the phase cancellation phenomenon is absent, so that such dips in reflectance do not appear, as is illustrated in Fig. 3b (with the two qubits switched compared to the current discussion).

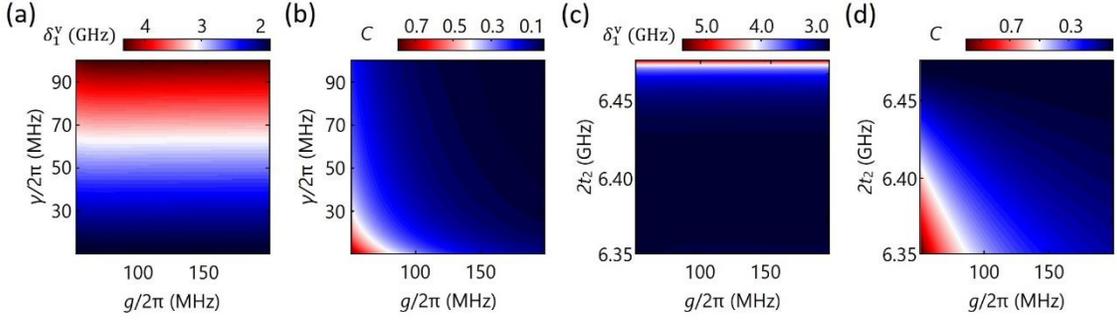

**FIG. S5.** Geometric features of the bright spots. (a, b) $\delta_1^V$ and $C$ as function of coherence rate $g = g_1^{\max} = g_2^{\max}$ and decoherence rate $\gamma = \gamma_1 = \gamma_2$, respectively. (c, d) $\delta_1^V$ and $C$ as a function of inter-dot tunnel coupling rate of DQD$_1$ coherence rate $g$ and $2t_2$.

## Section S7. **Qubit-qubit interaction in the frequency domain**

In the main text, we studied the qubit-qubit coupling by sweeping the qubit frequencies. For comparison, here we attempt to detect the interaction between the two qubits by varying the frequency of the probe microwave.

We first study the interaction between each individual DQD and the resonator. In Fig. S6a, we make DQD$_2$ far detuned from the resonator and fix DQD$_1$ tunnel coupling $2t_1$ near the resonator frequency $f_r = 6.48$ GHz. We then measure the microwave reflectance as a function of the probe frequency $\omega_p/2\pi$ and detuning $\delta_1$ of DQD$_1$ [4]. From the vacuum Rabi splitting (Fig. S6b), we can extract the parameters of $(g_1, \gamma_1)/2\pi = (81,65)$ MHz. Equivalent measurements are performed for DQD$_2$ indicating the parameters of $(g_2, \gamma_2)/2\pi = (80,55)$ MHz, as shown in Figs. S6c and d.

For a two-qubit study, we tune DQD$_2$ into resonance with the resonator and measure $|S_{11}|$ while scanning the detuning $\delta_1$ of DQD$_1$, as shown in Fig. S6e. Figure S6f shows that the enhanced Rabi splitting with the splitting $g_{\text{eff}}/2\pi = 112$ MHz $\approx \sqrt{g_1^2 + g_2^2}/2\pi = 113$ MHz, which is an evidence of qubit-qubit interaction.

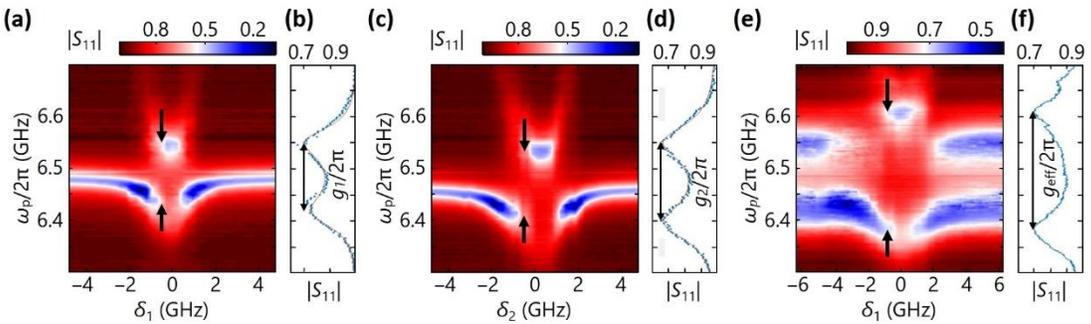

**FIG. S6.** Individual vacuum Rabi splitting and two-qubit enhance Rabi splitting. (a, c) Measured reflectance $|S_{11}|$ as a function of probe frequency $\omega_p/2\pi$ and detuning $\delta_{1(2)}$ of DQD$_{1(2)}$. (b, d) The

resonator spectrum $|S_{11}|(\omega_\mathrm{p})$ at $\omega_{\mathrm{a},1(2)}/2\pi = \omega_\mathrm{r}/2\pi$ (black arrow in (a) or (c)) (e) Measure reflectance $|S_{11}|$ as a function of $\omega_\mathrm{p}/2\pi$ and $\delta_1$ when qubit$_2$ is on resonance with the resonator. (f) Measured reflectance $|S_{11}|$ as a function of probe frequency $\omega_\mathrm{p}/2\pi$ and detuning $\delta_1$ when $\omega_{\mathrm{a},1}/2\pi = \omega_{\mathrm{a},2}/2\pi = \omega_\mathrm{r}/2\pi$ (black arrow in (e)).

## Section S8. **Resonator frequency change**

Our frequency-tunable SQUID array resonator is sensitive to the fluctuations of the electromagnetic environment. In our experiment we set the resonator working at its maximum frequency where it is insensitively to the first order of electromagnetic fluctuations. However, the resonator frequency may still shift during hours of continuous measurement. Here, we analysis the dependence of spectrum on the resonator detuning $i(\omega_\mathrm{r} - \omega_\mathrm{p})$. In Fig. S7, the solid lines are the calculated curves of $|S_{11}|$ along the diagonal $\delta_1 = \delta_2$ in Fig. 4a with different $(g, \gamma)$ from our estimated values. There are some discrepancies between the data and the calculated results based on the extracted parameters of $(g, \gamma)/2\pi = (80, 55)$ MHz at large detuning $\delta$, where we expect them to converge. Such differences could be caused by unwanted resonator detuning $i(\omega_\mathrm{r} - \omega_\mathrm{p})$ in Eq. (S1). Indeed, the simulated curve (black dashed line in Fig. S7) with a shifted resonator frequency $f_\mathrm{r}' = 6.4775$ GHz fits our date quite well. The frequency shift of the resonator of $\Delta f_\mathrm{r} = 2.5$ MHz is small, and our analysis is focused around the degenerate point (central regions in Fig. 3) where the deviations caused by the frequency shift are negligible. As such this frequency deviation will not affect our results in the main text, which is corroborated by the results in Fig. S8 with a new resonator frequency $f_\mathrm{r}' = 6.4775$ GHz (but otherwise with the same parameters as Figs. 3i-l). Our calculations in Fig. 4c and d have included the influence of the resonator frequency shift.

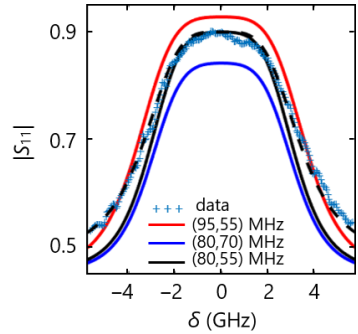

**FIG. S7.** $|S_{11}|$ as a function of $\delta_1 = \delta_2$. The sky-blue crosses are the data along the black dashed line in Fig. 3c. The red, blue and black lines are simulated with different values of $(g, \gamma)/2\pi$ at $2t_1 = 2t_2 = f_\mathrm{r} = 6.48$ GHz. The black dashed line is simulated with the parameters of $(g, \gamma)/2\pi = (80, 55)$ MHz and $(2t_1, 2t_2, f_\mathrm{r}', f_\mathrm{p}) = (6.48, 6.48, 6.4775, 6.48)$ GHz.

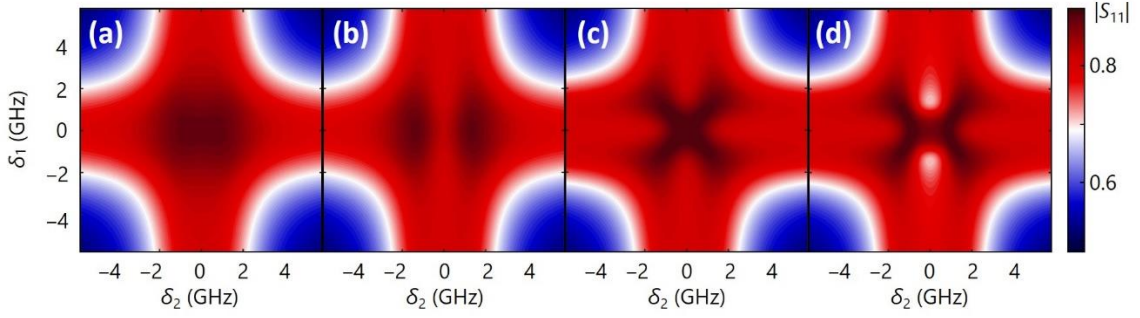

**FIG. S8.** Simulated correlated spectra $|S_{11}|$ as a function of $\delta_1$ and $\delta_2$ at $f_p = 6.48$ GHz and $f_r = 6.4775$ GHz using extracted parameters of $(g_1^{\max}, g_2^{\max}, \gamma_1, \gamma_2)/2\pi = (80, 80, 55, 55)$ MHz with (a) $(2t_1, 2t_2) = (6.68, 6.59)$ GHz, (b) $(2t_1, 2t_2) = (6.68, 6.48)$ GHz, (c) $(2t_1, 2t_2) = (6.48, 6.48)$ GHz, and (d) $(2t_1, 2t_2) = (6.48, 6.42)$ GHz.

## Section S9. **The application in spin qubit systems**

As discussed in the conclusion of the main text, our experimental approach of varying two physical parameters while keeping the probe frequency constant can be applied to other cavity QED coupled multi-qubit systems as well. Here we consider the example for spin qubits in the presence of a micromagnet.

One way to reach the strong coupling limit for a spin qubit is to introduce spin-charge hybridization [2,5-7], where the spin qubit in a DQD couples to the resonator photon mode via its electric field, mediated by the synthetic spin-orbit coupling produced by a micromagnet. The Hamiltonian of two such qubits, together with the cavity, is given by

$$H = \omega_r a^\dagger a + \sum_{i=1,2} \frac{1}{2}(\delta_i \sigma_z + 2t_i \sigma_z + B_z \sigma_z + B_x \sigma_x \tau_z) \cdot 2\pi + \sum_{i=1,2} g_{c,i}(\sigma_{+,i} a + a^\dagger \sigma_{-,i})$$

where $\sigma_{z(+,-)}$ and $\tau_z$ are the Pauli operators defined in charge qubit space and spin qubit space respectively, $a^\dagger$ ($a$) is the photon creation (annihilation) operator, $\omega_r$ is the resonator angular frequency, $\delta_i$ is the detuning of DQD$_i$, $2t_i$ is the interdot tunnel coupling of DQD$_i$, $B_{z(x)}$ is the magnetic fields in energy unit, $g_{c,i}$ is the charge-photon coupling strength at $\delta_i = 0$ and $\hbar = 1$.

The ground state and the first excited state (predominantly a pair of spin states at $B_z < 2t_i$) of the DQD is coded as $|0\rangle$ and $|1\rangle$ with the qubit frequency

$$\omega_{a,i}/2\pi = \frac{1}{2}\Big(\sqrt{\left(f_i + \sqrt{B_{z,i}^2 + B_{x,i}^2 \sin^2\theta_i}\right)^2 + B_{x,i}^2 \cos^2\theta_i}$$

$$-\sqrt{\left(f_i - \sqrt{B_{z,i}^2 + B_{x,i}^2 \sin^2\theta_i}\right)^2 + B_{x,i}^2 \cos^2\theta_i}\Big)$$

where $f_i = \sqrt{(2t_i)^2 + \delta_i^2}$ is the charge qubit energy splitting and $\theta_i = \arctan\frac{\delta_i}{2t_i}$ is the 'orbital mixing angle'. The spin-photon coupling strength is given by $g_{s,i} = g_{c,i}\cos\theta_i\sin\phi_i$, where $\phi_i =$ $\arctan \frac{B_{x,i}\cos\theta_i}{f_i - B_{z,i}}$ is the spin-orbit mixing angle determined by the field gradient $B_{x,i}$ [6].

With spin qubits coupled to the resonator via the DQD charge qubit, any modification of the charge qubit parameters would change spin-photon coupling accordingly. The input-output theory dictates that the microwave response is modulated by the spin qubits, yielding a microwave reflectance of

$$S_{11} = -1 + \frac{\kappa_{\text{ext}}}{i(\omega_{\text{r}} - \omega_{\text{p}}) + g_{s,1}\chi_1 + g_{s,2}\chi_2 + \kappa_{\text{tot}}/2}$$

where $\omega_p$ is the probe angle frequency, $\chi_i = \frac{g_{s,i}}{i(\omega_{a,i} - \omega_p) + \gamma_{s,i}}$ is the electric susceptibility, $\gamma_{s,i} = \gamma_{s,i}(\gamma_{c,i}, 2t_i, \delta_i, B_x, B_z)$ is the effective decoherence rate of spin qubit, $\gamma_c$ is the charge decoherence rate, $\kappa_{\text{ext}}$ and $\kappa_{\text{tot}}$ are the external loss and total loss of the resonator, respectively. Here, to make the content expressed by the equation more intuitive, we have omitted the terms of higher energy levels since these terms have little influence on the microwave response when $B_z < 2t_i$.

In Fig. S9, by tuning tunnel coupling $t_i$ of each double quantum dot, we simulate the correlated spin spectrum using the parameters of $(g_{c,1}, g_{c,2}, \gamma_{c,1}, \gamma_{c,2})/2\pi = (40, 40, 85, 85)$ MHz, $B_z = B_{z,1} = B_{z,2} = 6.856$ GHz, $B_x = B_{x,1} = B_{x,2} = 1.028$ GHz, and $f_r = 6.745$ GHz extracted from Ref. [2]. The simulated spectra contain similar geometrical patterns shown in Figs. 3e-h, and we can readily identify coupling regimes for both spin qubits.

Notice that in this scheme spin states are coupled with the resonator through its electric field via spin-charge hybridization. As such, even though the patterns in Fig. S9 are obtained by varying the charge qubit parameters (specifically interdot detunings), it reflects the low energy spectrum corresponding to spin states and couplings.

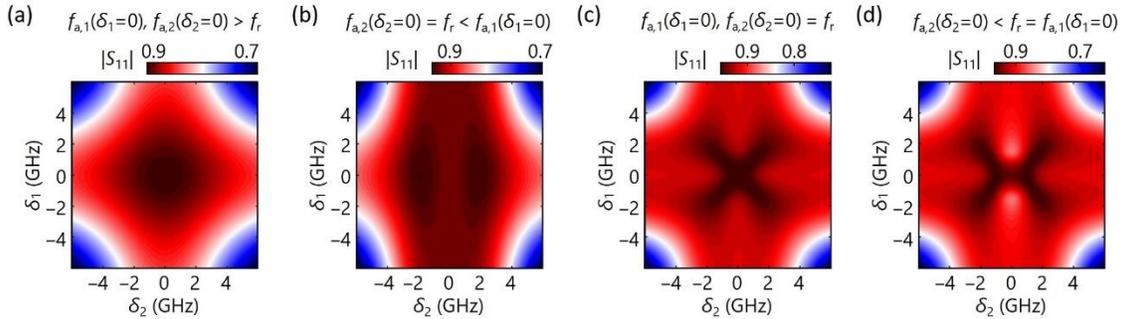

**FIG. S9.** Simulated two-spin-qubit spectra as a function of $\delta_1$ and $\delta_2$ at $f_p = f_r = 6.745$ GHz and $(B_z, B_x) = (6.856, 1.028)$ GHz with (a) $(2t_1, 2t_2) = (10, 10)$ GHz, (b) $(2t_1, 2t_2) = (10, 8.8)$ GHz, (c) $(2t_1, 2t_2) = (8.8, 8.8)$ GHz, (d) $(2t_1, 2t_2) = (8.8, 8.7)$ GHz.

## Section S10. **Searching for qubit resonance point**

In the main text, we point out that with our experimental approach we can search for multi-qubit resonance points through pattern recognition. Here we give a comparison between our approach and the conventional frequency sweeping technique. Specifically, we simulate the enhanced vacuum Rabi splitting and the correlated spectra of two coupled qubits mediated by microwave photons when $2t_1 = f_r = 6.48$ GHz $\sim 2t_2$. In the frequency domain, Figs. S10a-c present reflection data for three slightly different $t_2$. The subtle changes among the three panels are difficult to discern, so that it is hard to identify whether $2t_1 = 2t_2$. As a comparison, our spectroscopy method provides clearly evolving features, such as the bright spots in Fig. S10d, the X-like pattern in Fig. S10e and the fading and the splitting of the central area in Fig. S10f, for $2t_2 = 6.46$ GHz, $6.48$ GHz and $6.50$ GHz, respectively. These features could help us identify where the resonance point is more quickly and decisively.

We have also simulated the enhanced vacuum Rabi splitting and the correlated spectra of three coupled qubits mediated by microwave photons when $2t_1 = 2t_2 = f_r = 6.48$ GHz $\sim 2t_3$, where $t_i$ is tunnel coupling of qubit$_i$. In the frequency domain, it is again difficult to catch the differences among Figs. S11a-c, and pick out the case $2t_1 = 2t_2 = 2t_3$. In comparison, Figs. S11d-f provide three distinctive patterns, and demonstrate that our method can be a quick and intuitive way to resolve collective properties of multiple qubits.

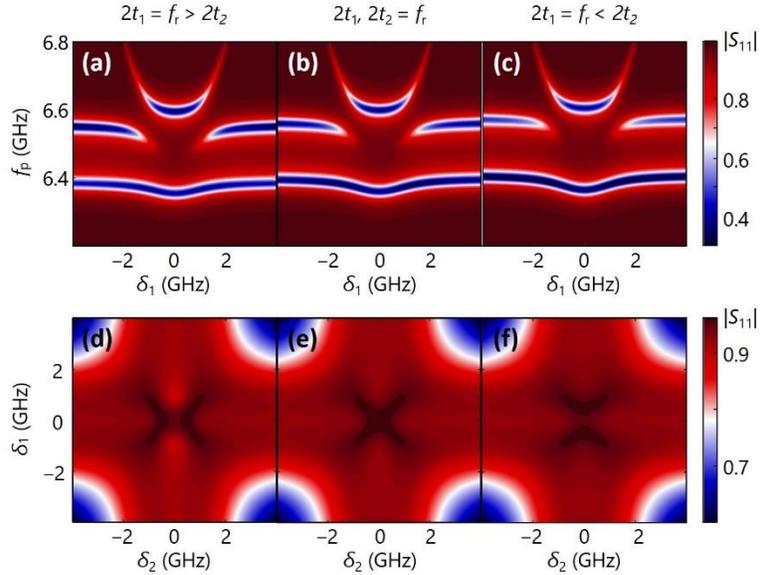

**FIG. S10.** Simulated enhanced vacuum Rabi splitting and correlated spectra of two coupled qubits mediated by microwave photons. The enhanced vacuum Rabi splitting in the frequency domain as a function of $\delta_1$ and $\omega_p/2\pi$ at (a) $2t_2 = 6.46$ GHz, (b) $2t_2 = 6.48$ GHz and (c) $2t_2 = 6.50$ GHz. (d-f) show the corresponding two-qubit spectra as a function of $\delta_1$ and $\delta_2$ at $f_p = f_r = 6.48$ GHz.

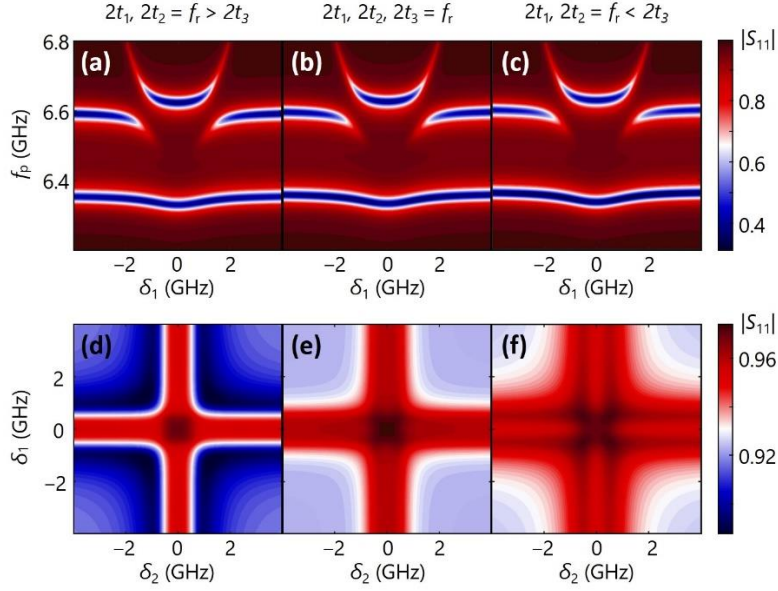

**FIG. S11.** Simulated enhanced vacuum Rabi splitting and correlated spectra of three coupled qubits mediated by microwave photons. The enhanced vacuum Rabi splitting in the frequency domain as a function of $\delta_1$ and $\omega_p/2\pi$ at (a) $2t_3 = 6.46$ GHz, (b) $2t_3 = 6.48$ GHz and (c) $2t_3 = 6.50$ GHz. (d-f) show the corresponding two-qubit spectra as a function of $\delta_1$ and $\delta_2$ at $f_p = f_r = 6.48$ GHz.

## Section S11. **Two-qubit spectra with different $g_i$ and $\gamma_i$**

In the main text, we assume that decoherence rates and qubit-resonator coupling strengths of the two qubits are the same. However, our method is a quite general and can be applied to cases when $g_1^{max}$ and $g_2^{max}$ ($\gamma_1$ and $\gamma_2$) are different. To emphasize this point, here we simulate the spectra using different $g_i^{max}$ and $\gamma_i$.

In Fig. S12, the numerical results show the same qualitative behavior when $g_1^{max}$ and $g_2^{max}$ are different. The geometric patterns in the spectra are still highly distinctive and unique. Comparing the three panels in Fig. S12, we find that the increased $g_1^{max}$ enhances qubit$_1$'s contribution in the microwave response and stretch out the spectra pattern along the Y-axis. In Fig. S13, the two-qubit spectra are simulated with different $\gamma_1$ and $\gamma_2$, again showing the similar results. The decreased $\gamma_1$ enhances qubit$_1$'s contribution in the microwave response, but does not change the qualitative behavior of the overall response. As for the case $(2t_1, 2t_2) = (6.48, 6.48)$ GHz shown in Figs. S13c, g and k, the X-like pattern is stretched along the Y-axis as $\gamma_1$ increases. As such, from these distinctive patterns, the parameter regimes for both qubits can still be readily identified.

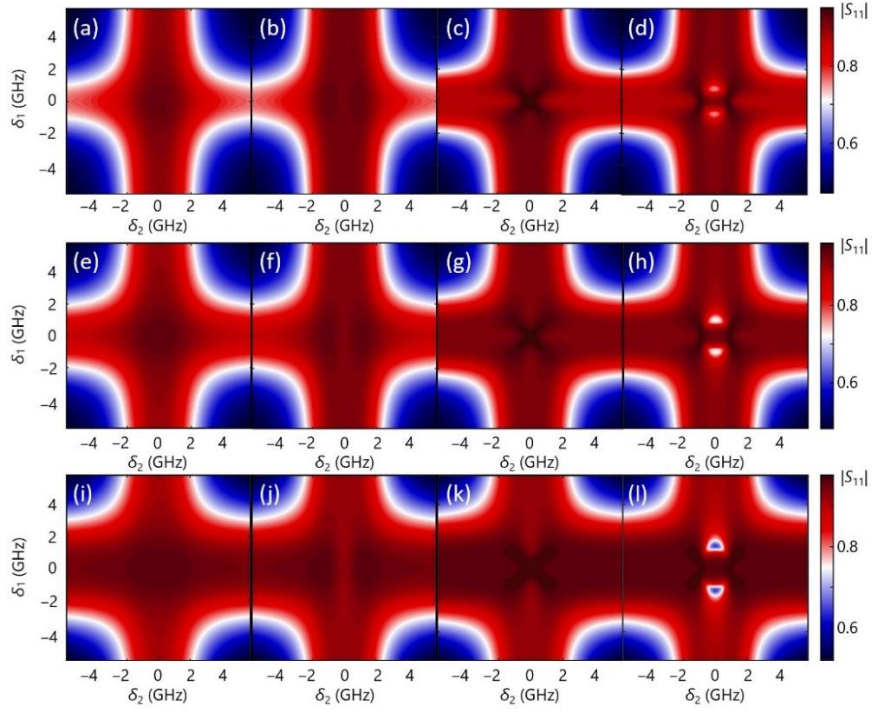

**FIG. S12.** Calculated two-qubit spectra as a function of $\delta_1$ and $\delta_2$ at $f_p = f_r = 6.48$ GHz, with (a) $(2t_1, 2t_2) = (6.68, 6.59)$ GHz, (b) $(2t_1, 2t_2) = (6.68, 6.48)$ GHz, (c) $(2t_1, 2t_2) = (6.48, 6.48)$ GHz, and (d) $(2t_1, 2t_2) = (6.48, 6.42)$ GHz. The parameters in (a-d) are $(g_1^{\max}, g_2^{\max}, \gamma_1, \gamma_2)/2\pi = (65, 85, 22, 22)$ MHZ. Simulations corresponding to (a-d) with the parameters $(g_1^{\max}, g_2^{\max}, \gamma_1, \gamma_2)/2\pi = (85, 85, 22, 22)$ MHz and $(g_1^{\max}, g_2^{\max}, \gamma_1, \gamma_2)/2\pi = (125, 85, 22, 22)$ MHz are shown in (e-h) and (i-l) respectively.

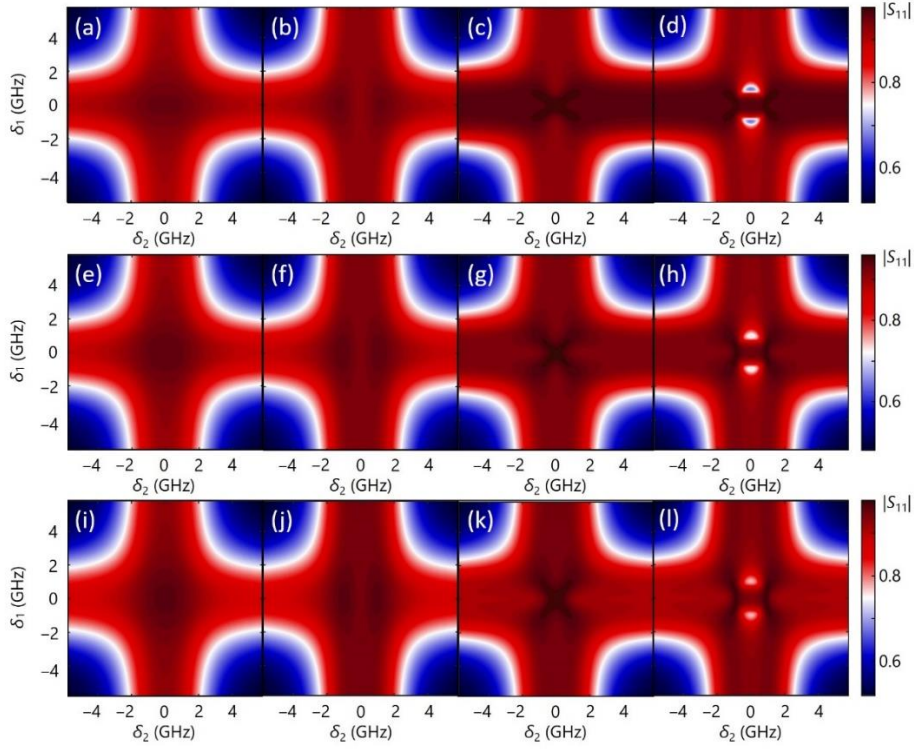

**FIG. S13.** Calculated two-qubit spectra as a function of $\delta_1$ and $\delta_2$ at $f_\mathrm{p} = f_\mathrm{r} = 6.48$ GHz, with (a) $(2t_1, 2t_2) = (6.68, 6.59)$ GHz, (b) $(2t_1, 2t_2) = (6.68, 6.48)$ GHz, (c) $(2t_1, 2t_2) = (6.48, 6.48)$ GHz, and (d) $(2t_1, 2t_2) = (6.48, 6.42)$ GHz. The parameters in (a-d) are $(g_1^{\max}, g_2^{\max}, \gamma_1, \gamma_2)/2\pi = (85, 85, 11, 22)$ MHZ. Simulations corresponding to (a-d) with the parameters $(g_1^{\max}, g_2^{\max}, \gamma_1, \gamma_2)/2\pi = (85, 85, 22, 22)$ MHz and $(g_1^{\max}, g_2^{\max}, \gamma_1, \gamma_2)/2\pi = (85, 85, 44, 22)$ MHz are shown in (e-h) and (i-l) respectively.